\documentclass[10pt,twocolumn,letterpaper]{article}

\usepackage{wacv}              

\usepackage{graphicx}
\usepackage{amsmath}
\usepackage{amssymb}
\usepackage{booktabs}

\usepackage{algorithmic}
\usepackage{textcomp}
\usepackage{xcolor}
\usepackage{flushend}
\usepackage{tabularx}
\usepackage{multirow}
\usepackage{booktabs,enumitem}
\usepackage[linesnumbered,ruled,lined,commentsnumbered]{algorithm2e}
\usepackage{color,soul}
\usepackage[pagebackref,breaklinks,colorlinks]{hyperref}

\usepackage[capitalize]{cleveref}
\crefname{section}{Sec.}{Secs.}
\Crefname{section}{Section}{Sections}
\Crefname{table}{Table}{Tables}
\crefname{table}{Tab.}{Tabs.}

\begin{document}

\title{RankDVQA: Deep VQA based on Ranking-inspired Hybrid Training}

\author{Chen Feng, Duolikun Danier, Fan Zhang and David Bull\\
\textit{Visual Information Laboratory, University of Bristol, Bristol, UK, BS1 5DD}\\
{\{chen.feng, duolikun.danier, fan.zhang, dave.bull\}@bristol.ac.uk}}

\maketitle

\begin{abstract}
   In recent years, deep learning techniques have shown significant potential for improving video quality assessment (VQA), achieving higher correlation with subjective opinions compared to conventional approaches. However, the development of deep VQA methods has been constrained by the limited availability of large-scale training databases and ineffective training methodologies. As a result, it is difficult for deep VQA approaches to achieve consistently superior performance and model generalization. In this context, this paper proposes new VQA methods based on a two-stage training methodology which motivates us to develop a large-scale VQA training database without employing human subjects to provide ground truth labels. This method was used to train a new transformer-based network architecture, exploiting quality ranking of different distorted sequences rather than minimizing the difference from the ground-truth quality labels. The resulting deep VQA methods (for both full reference and no reference scenarios), FR- and NR-RankDVQA, exhibit consistently higher correlation with perceptual quality compared to the state-of-the-art conventional and deep VQA methods, with average SROCC values of 0.8972 (FR) and 0.7791 (NR) over eight test sets without performing cross-validation. The source code of the proposed quality metrics and the large training database are available at \url{https://chenfeng-bristol.github.io/RankDVQA}.
\end{abstract}
\section{Introduction}
\label{sec:intro}

With the explosion of video streaming and conferencing services, there has been a surge in the prevalence of video content on the internet. It is reported that videos account for 82\% of the global internet traffic, with 4.8 Zetabytes of video data being transmitted annually~\cite{r:cisco2020}. Given such increasing demand, it is important that the quality of the transmitted videos matches the requirements of the service provider and the expectations of the end-user. To this end, video quality assessment (VQA) methods are employed to provide an objective measure of the perceived video quality. They therefore represent a crucial component in the video coding process, enabling operating points exhibiting appropriate trade offs between video quality and bit rate. Other than video compression, VQA methods are also used to assess the performance of various  processing tasks such as denoising \cite{liu2010high,tassano2020fastdvdnet}, restoration \cite{liang2022vrt,wang2019edvr}, frame interpolation \cite{FIoLPIPS, hou2022perceptual} and super-resolution \cite{lucas2019generative, ma2019perceptually}.

\begin{figure}[tbp]
\centering
\hspace{-0.5cm}\includegraphics[width=1\linewidth]{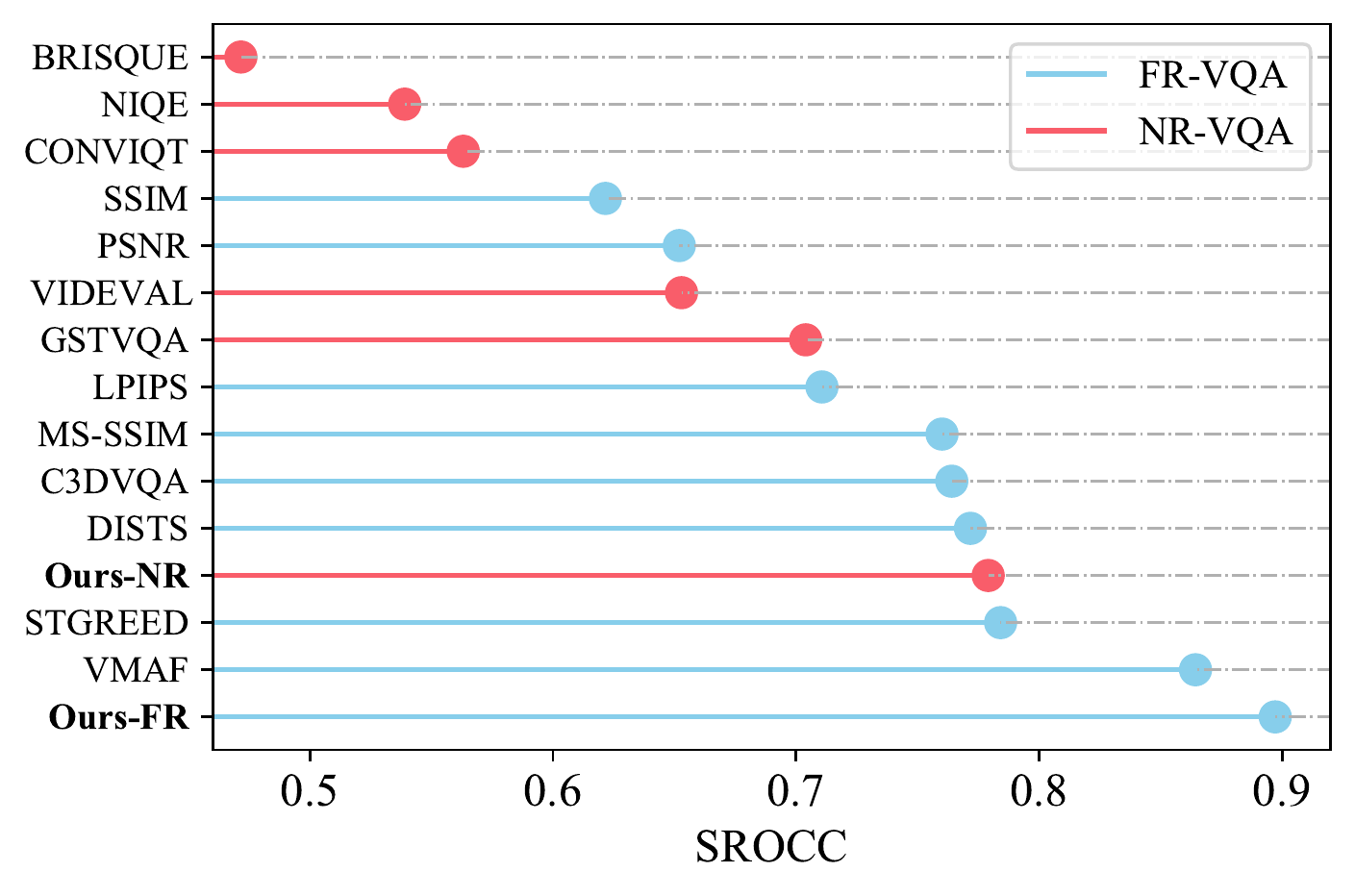}

\caption{The performance of the proposed FR- and NR-RankDVQA and selected benchmark VQA methods. Here the average SROCC values were used as the statistical parameter to measure their correlation with subjective ground truth.  Both FR- and NR-RankDVQA achieve the best performance in each category (FR and NR), and NR-RankDVQA even outperforms some FR quality metrics. The results for all the deep VQA methods shown here are based on the fixed model parameters without performing re-training  on each database (intra database cross-validation).
\label{fig:Bar}}
\vspace{-10pt}
\end{figure}

Conventional VQA approaches rely on signal processing techniques to measure distortions. Notable examples include PSNR, SSIM \cite{ssim} and its variants \cite{c:mssim,ssimplus,vssim}, VIF \cite{VIF} and VIIDEO \cite{VIIDEO}, VBLIINDS \cite{VBLIINDS}. With the increasing popularity of learning-based techniques, the hand-crafted features in these metrics have also been enhanced using machine learning models such as Support Vector Regressors to achieve improved prediction performance, e.g., VMAF~\cite{w:VMAF}. More recently, deep learning has driven the development of video quality assessment methods \cite{Kim2018DeepVQ,C3DVQA},  achieving promising results when compared to classical approaches. However, deep VQA methods tend to be constrained by the following issues. (i) \textbf{The lack of reliable large and diverse training databases}: Most existing methods were trained using a relatively small video database (typically with only a few hundreds subjective labels)\footnote{Note that different VQA databases cannot be simply combined because of different experimental settings during subjective data collection.}. This is widely acknowledged to be insufficient for training a model with a relatively high network capacity. This is why most existing deep VQA methods only perform well when cross-validated on a single database yet show unsatisfactory cross-dataset generalization ability, with inconsistent overall performance compared to conventional or regression-based VQA methods, such as VMAF \cite{w:VMAF}. (ii) \textbf{Ineffective training methods}: The training processes used were designed to consider one distorted content at a time. In contrast, the ability of a metric to differentiate the quality of differently distorted versions of the same (or different) source content is an important characteristic of a VQA method which has not yet been exploited.

To address the above issues, in this paper, we propose a ranking-based hybrid training methodology for deep VQA. The main contributions of this work are summarized below.

\begin{itemize}[leftmargin=*]
\setlength\itemsep{0em}
    \item This paper proposes a \textbf{new two-stage training methodology}, which combines patch-wise VQA with spatio-temporal pooling. This first uses VMAF-based quality ranking information to train a deep VQA model, and then employs multiple small video databases with subjective ground truth to train an aggregation network for spatio-temporal pooling. By using this new training framework, the proposed deep VQA methods can achieve significantly improved model generalization, and avoid the need to perform intra-database cross-validation which is not a  practical evaluation method for real-world applications.
      
    \item More importantly, by using VMAF to generate ranking-based labels, we have, for the {first} time, developed a \textbf{large-scale training dataset} (204,800 video dodecuplet groups with VMAF ranking-based annotations) for optimizing patch-level VQA models (Stage 1) without performing costly subjective tests. 
    This solution circumvents problem (i)  - the lack of reliable large training databases.
    
     

     \item During the training processes in both stages, we utilize the \textbf{quality ranking information associated with different distorted videos} (from the same or different source content). As such, we reformulate the primary aim of VQA as differentiating the quality of two distorted videos, rather than providing an absolute index for an individual sequence.
    This takes account of issue (ii) - ineffective training methods, and for the first time enables the combination of multiple VQA databases for training (in Stage 2). To the best of our knowledge, we are the {first} to adopt ranking-based training for VQA and extend it by additionally considering dual-source ranking information. 

    \item Based on the new training methodology, we have trained a full reference (FR) and a no-reference (NR) deep VQA method (RankDVQA), employing a transformer-based architecture (for Stage 1) and a CNN-based aggregation network (for Stage 2). Through comprehensive evaluation on eight commonly used video quality datasets, we show that both proposed metrics (FR-and NR-RankDVQA) outperform their competing approaches including VMAF (for the full reference metric) without re-training or cross-validation on each database. This is shown in Figure \ref{fig:Bar}, where the average SROCC values are 0.8972 (FR) and 0.7791 (NR). To the best of our knowledge, the proposed FR quality metric is the first that consistently outperforms VMAF on various video quality databases without re-training.
\end{itemize}

\section{Related Work}
\label{sec:related work}

Depending on the availability of reference content, VQA methods can be classified into three sub-groups: full reference (FR), reduced reference (RR) and no reference (NR). RR and NR models require either partial or no information about the reference video, while FR approaches use both the reference and the distorted video content as inputs. The work mainly focuses on the FR and NR scenarios.

\begin{figure*}[htbp]
\centering
\includegraphics[width=0.95\linewidth]{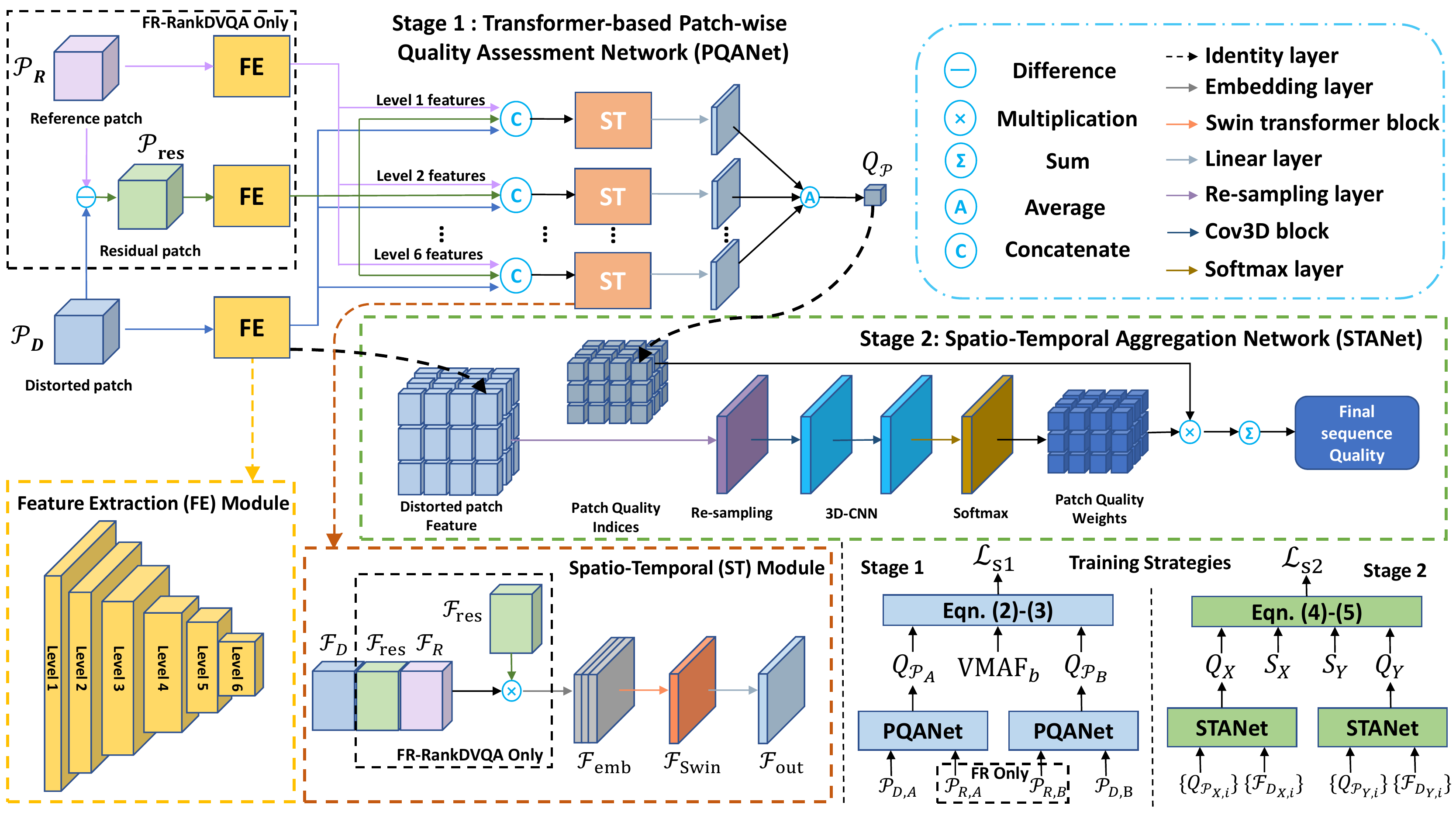}
\caption{Illustration of the proposed approach including network architectures and training strategies in both stages. \label{fig:framework}
}
\end{figure*}

\noindent\textbf{FR VQA methods.} Among existing FR VQA methods, PSNR and SSIM \cite{ssim} are the most widely used approaches, primarily due to their simplicity and low computational complexity. However, to achieve better correlation with subjective opinions, many perceptually-optimized quality metrics have been proposed. These include  SSIM variants \cite{c:mssim,ssimplus,vssim}, MAD \cite{MAD}, STMAD \cite{stmad}, MOVIE \cite{MOVIE} and PVM \cite{PPVM}. Netflix has developed a machine learning based quality metric, VMAF \cite{w:VMAF}, combining six different video features using a Support Vector Regressor (SVR), which provides consistently better correlation  with subjective opinions over a wide range of content and distortion types than most conventional VQA methods. 

In recent years, deep neural networks have been increasingly applied to image \cite{8063957,Kim_2017_CVPR,Ahn_2021_CVPR} and video \cite{Kim2018DeepVQ,C3DVQA,E2E,CONTRIQUE,STGREED,DISTS,wu2022discovqa} quality assessment. Such methods have demonstrated the potential to compete with conventional quality metrics. However, due to the limited cardinality of available subjective databases, most deep VQA methods have been trained and evaluated on small databases through cross validation. To train Convolutional Neural Network (CNN) models,  each video sequence is typically segmented into a number of patches, all of which are labeled with the sequence-level subjective quality scores. This sub-optimal solution often results in overfitting and inconsistent performance across different databases without re-training.

\noindent\textbf{NR VQA methods.} NR VQA models estimate the quality of a distorted video without any reference information. Similar to FR methods, conventional NR quality metrics usually extract features from the distorted video and compare them in various (spatial, temporal, spatio-temporal, and frequency and wavelet) domains. Notable examples include BRISQUE \cite{brisque}, NIQE \cite{niqe}, VIIDEO \cite{VIIDEO}, Video BLIINDS \cite{VBLIINDS}, BIQI \cite{biqi} and DIIVINE \cite{DIIVINE}.  Recently, a number of deep learning-based NR VQA methods, including DeepVBQA \cite{DeepVBQA}, TLVQM \cite{CNNTLVQM}, DeepSTQ \cite{zhou2020deep}, VSFA \cite{VSFA}, RAPIQUE \cite{tu2021rapique}, MDTVSFA \cite{MDTVSFA}, VIDEVAL \cite{VIDEVAL} and GSTVQA \cite{GSTVQA} have been proposed, providing enhanced performance over classic methods, although their development is still constrained with the same issue with deep FR methods - lack of sufficient and diverse training material.

\noindent\textbf{Learning to rank.} Most deep VQA methods mentioned above were trained by minimizing the $\ell 1$/$\ell 2$ distance to the subjectively assessed ground truth. However, it is more important for a quality metric to provide accurate ranking results given different distorted sequences. To this end, learning-to-rank based methods have been proposed in the context of no reference image quality assessment \cite{LRBIQA,Tres,rankiqa,dipIQ,DBCBIQA,UABIQA}. However, these methods do not fully exploit the ranking information associated with distorted versions generated from both single and multiple sources. Furthermore, this approach has not been fully applied for generic VQA - it has only been used within a bespoke video frame interpolation quality metric \cite{hou2022perceptual}.

\section{Proposed Method: RankDVQA}
\label{sec:Proposed Approach}

The architecture of RankDVQA is shown in Figure \ref{fig:framework}. The input distorted video sequence (together with its corresponding original counterpart for the FR case) is first segmented into non-overlapped $H\times W\times C\times N$ video volume (patches). Here $H$ and $W$ correspond to the spatial resolution of the each image, and they are set to 256. $C=3$ is the number of channels (YCbCr) and $N$ is the number of sampled frames, set to 12 here, following \cite{hou2022perceptual}. The input dodecuplet(s) is first processed by a transformer-based network which produces a patch-level quality index (Stage 1). The indices for all the dodecuplets in the same sequence will be passed to an aggregation network that outputs the final sequence level quality index (Stage 2). The network architecture, training material and methodologies used in these two stages are described below.

\subsection{Stage 1: Patch Level Quality Assessment}
\label{sec:stage1}

\noindent\textbf{Network architecture.} In the first stage, the proposed method employs a transformer-based patch quality assessment network, PQANet, to predict a quality index, given the input distorted patch, $\mathcal{P}_D$, (and its reference counterpart, $\mathcal{P}_R$, in the full reference scenario), with a size of 256$\times$256$\times$3$\times$12. If this network is used for FR quality assessment, a residual patch, $\mathcal{P}_\mathrm{res}$ is also calculated\cite{Kim_2017_CVPR,Kim2018DeepVQ},
\begin{equation}
\mathcal{P}_\mathrm{res} = \frac{\log(1/(\mathcal{(\mathcal{P}_R - \mathcal{P}_D) }^2 + \mathcal{E}/(2^B-1)^2)}{\log((2^B-1)^2/\mathcal{E})},
\label{eq:Res}
\end{equation}
where $\mathcal{E}$ equals 1 \cite{Kim_2017_CVPR,Kim2018DeepVQ}, and $B$ is the video bit depth.

The network architecture used in the first stage (shown in Figure \ref{fig:framework}) is similar to that reported in \cite{hou2022perceptual}, which is based on Swin Transformer \cite{liu2021swin} blocks. Similar designs have also been used in \cite{liu2021swin, wu2022fast, jiang2022self}. This network first employs a Feature Extraction (FE) module (a pyramid network) to extract six levels of feature maps for each input patch. Each level in the pyramid network comprises two 3 $\times$ 3 Conv2D layers, with the second layer having a stride of 2 for downsampling. The number of channels in each level increases in a progressive manner, starting with 16 in the first level and reaching 128 in the final level. For each level, we concatenate the features from all input patches to obtain the input for the spatio-temporal (ST) module.

Based on the concatenated feature maps at each level, we employ the ST module to estimate the spatio-temporal quality at this level. The ST module first normalizes all input features in the channel dimension  \cite{zhang2018unreasonable}, and concatenates them together. Different from \cite{hou2022perceptual}, in the full reference scenario, we perform an element-wise multiplication between residual features, $\mathcal{F}_\mathrm{res}$, and the concatenated features, $[\mathcal{F}_D \ \mathcal{F}_\mathrm{res}\  \mathcal{F}_R]$, (where $\mathcal{F}_D, \mathcal{F}_R$ are extracted from $\mathcal{P}_D$ and $\mathcal{P}_R$), before feeding them into the linear embedding layer that projects features to a fixed dimension of 32. This has been proved to offer evident improvement over absolute position embedding in \cite{liu2021swin}. The output, $\mathcal{F}_\mathrm{emb}$ is then processed through a Swin Transformer block \cite{liu2021swin} and a linear layer to obtain the quality index at this feature level. Finally the indices for all six feature levels are averaged to calculate the final patch level quality index, $Q_\mathcal{P}$.



\noindent\textbf{Training Database Generation.} As highlighted in Section 1, there are two primary issues with most current deep VQA methods, (i) a lack of large diverse training databases (ii) ineffective training strategies. To overcome the first issue, instead of using small video databases containing subjective opinion scores, we developed a large-scale training database containing diverse distorted sequences with artifacts commonly encountered in modern video streaming scenarios. 

To generate the training content, we selected 230 source sequences from the BVI-DVC dataset \cite{ma2021bvi} and the training database used in the CVPR 2022 CLIC video compression challenge \cite{clic}. Each source sequence was segmented into 64 frames and converted to YCbCr 4:2:0 format. These sequences were then compressed using four standard video codecs: H.264/AVC x264 \cite{merritt2006x264}, H.265/HEVC HM 16.20 \cite{r:HEVC}, AOM/AV1 1.0.0 \cite{w:AV1} and H.266/VVC VTM 7.0 \cite{s:VVC1}, at four quantisation levels to create diverse distortion types. The codec versions and configurations are summarized in \textit{Supplementary Material}. To further augment the training data, we also generated training content through resolution adaptation, which is commonly used in video streaming servers \cite{w:DO}. Specifically, all the source sequences were first down-sampled to three lower spatial resolutions, by a factor of 1.5, 2 or 3, using the Lanczos3 filter \cite{Lanczos3}. These low resolution sequences were then compressed by the four codecs mentioned above with the same coding configurations (also with four quantisation levels). Each compressed low resolution video was then decoded and up-sampled to its original resolution using the Lanczos3 filter to obtain the distorted sequence. This results in a total number of 14720 (230 sources $\times$ 4 resolutions $\times$ 4 codecs $\times$ 4 quantization levels) distorted sequences. All the distorted videos are converted to YCbCr 4:4:4 format for patch generation.

In order to perform ranking-based training, we produce patch level training material based on two approaches - single-source patch generation and dual-source patch generation. The former is used to improve the ability of the employed network to identify the quality difference between distorted versions of the same source. This is an important property of a VQA method when it is used for comparing the quality of videos generated by different methods, such as for video codec comparison. On the other hand, training on dual-source patches allows the model to produce more reliable quality scores for videos with different content characteristics. We note that these two characteristics have not been previously exploited in the development (or training) of VQA methods.

To produce single-source patches, we first randomly selected two distorted sequences corresponding to the same reference sequence, then randomly segmented each distorted sequence and its reference counterpart using a non-overlapping spatio-temporal sliding window of size 256$\times$256$\times$12. Each segment generates three dodecuplets (one for each distorted version, and one for the reference). Secondly, to generate dual-source patches, we randomly selected two distorted sequences, and segmented each (and their respective reference videos) using the same sliding window as mentioned above. The resulting patches can come from different source sequences, or from the same source sequence but at different spatio-temporal locations.  Doing this for each segment produces four dodecuplets (one for each distorted patch, and one for each reference patch). It is noted that, for training our NR VQA method, only distorted patches are used, while both distorted and reference patches are employed for the training of the full reference quality metric, as shown in Figure \ref{fig:framework}.

To label all these patches with reliable quality scores which correlate well with subjective ground truth, inspired by the previous contributions in \cite{ProxIQABovik,2021differentiable,DEMI}, we used the VMAF (Video Multi-method Assessment Fusion) metric to produce training targets for generated patches by comparing each distorted patch with its corresponding reference. It is further noted that, although VMAF provides relatively consistent correlation with perceptual quality on various video quality databases, its average correlation coefficient (SROCC) of 0.85 is far from perfect \cite{2021zhangPCS}. 

\begin{figure}[t]
\centering
\begin{minipage}[b]{0.495\linewidth}
\includegraphics[width=1.1\linewidth]{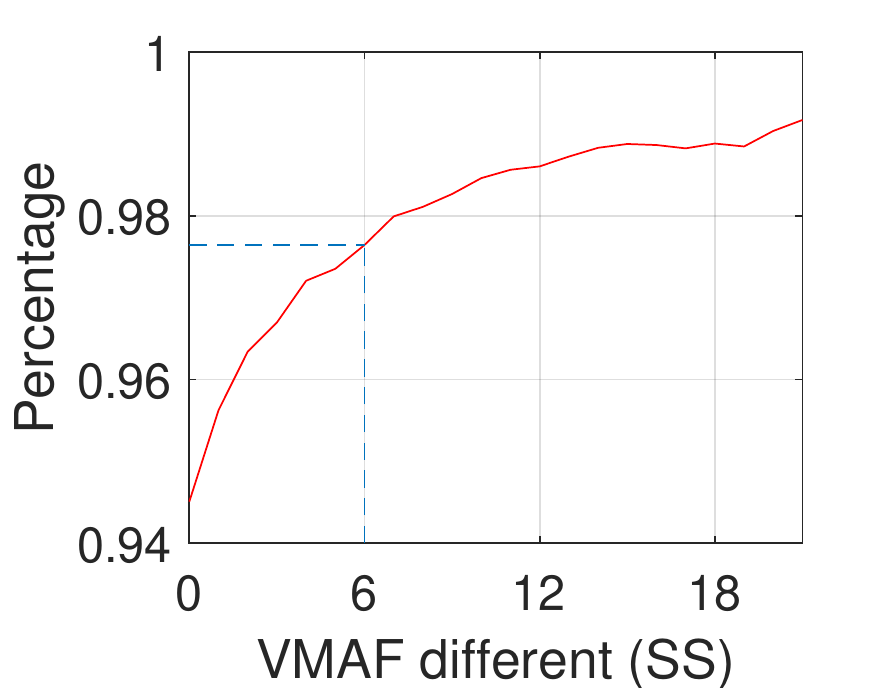}
\end{minipage}
\begin{minipage}[b]{0.495\linewidth}
\includegraphics[width=1.1\linewidth]{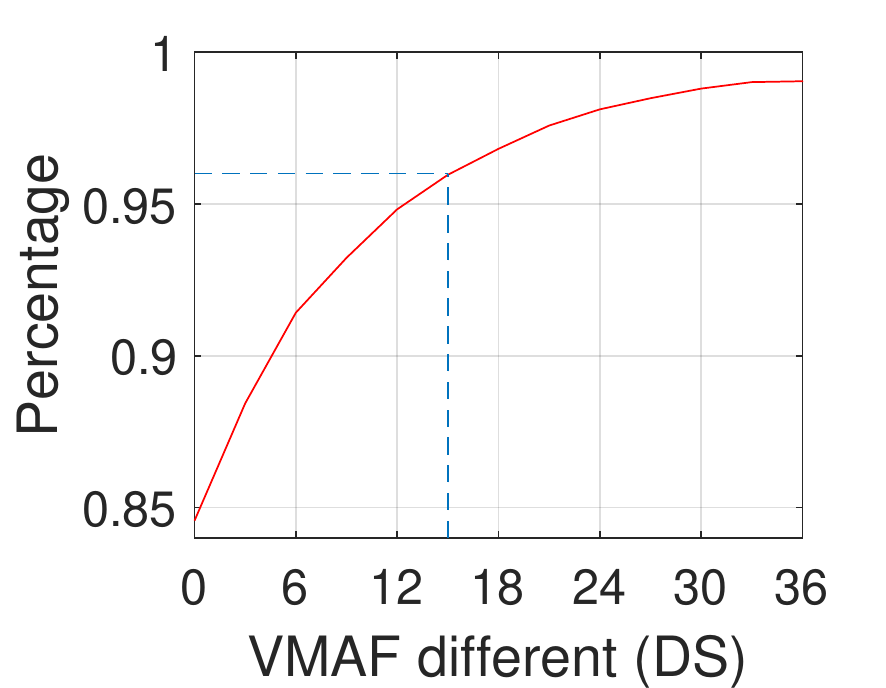}
\end{minipage}
\caption{VMAF difference between every two distorted sequences versus the accuracy of VMAF compared to subjective ground truth on the VMAF+ \cite{w:VMAF} database. (Left) Single sources. (SS) (Right) Dual sources (DS). }
\label{fig:vmafdiff}
\vspace{-10pt}
\end{figure}

To further improve the reliability of the training material, we adopted a similar approach to that in \cite{hou2022perceptual} to evaluate the performance of VMAF on an existing database, VMAF+ \cite{w:VMAF}, in terms of the accuracy to differentiate the quality of different distorted sequences of the same (or different) source sequences. The results are summarized in Figure \ref{fig:vmafdiff}. It can be observed that, when the VMAF value difference (between two distorted videos) is larger than 6 (for single source) or 15 (for dual sources), the differentiation accuracy according to the subjective score difference is above 95\%\footnote{We have used the same accuracy ratio 97.5\% as in \cite{hou2022perceptual,rankiqa} for single source cases. A slightly lower ratio (96\%) was utilized for dual source cases in order to remove fewer outliers (to obtain more training samples).}. We therefore use these two thresholds to remove potential outliers in the labeled patches. Specifically, for patches in each segment, if the absolute VMAF value difference between two distorted versions is smaller than the threshold (for single source or dual source), we exclude this segment from the final training dataset. In total, this generated approximately 204,800 groups (each of them corresponds to a single segment) of dodecuplet patches associated with the same number of VMAF difference labels, in which there are 30\% single sourcing patches and 70\% dual sourcing patches.

\noindent\textbf{Ranking-inspired Training Strategy}
As shown in Figure \ref{fig:framework}, the employed PQANet is trained in a Siamese method similar to that in \cite{yi2014deep,wu2018and,rankiqa}. Given two input distorted patches $\mathcal{P}_{D,A}$, $\mathcal{P}_{D,B}$ (and their corresponding reference patches $\mathcal{P}_{R,A}$, $\mathcal{P}_{R,B}$ for FR VQA) from the same patch group, the  difference between outputs of the patch quality assessment network, $Q_{\mathcal{P}_A}$ and $Q_{\mathcal{P}_{B}}$, are compared against their associated VMAF difference (after binarization), $\mathrm{VMAF}_b$ ($\mathrm{VMAF}_b=1$ if the VMAF value of patch A is larger than patch B. Otherwise, $\mathrm{VMAF}_b=0$). Specifically, we first calculate a probability $p$ with a sigmoid layer,
\begin{equation}
\begin{split}
{p = \mathrm{sigmoid}(Q_{\mathcal{P}_A} - Q_{\mathcal{P}_B})},
\label{eq:sigmoid}
\end{split}
\end{equation}
and obtain the binary cross entropy loss as:
\begin{equation}
\mathcal{L}_{s1} = - (\mathrm{VMAF}_b\log(p) + (1 -\mathrm{VMAF}_b )\log(1-p)).
\label{eq:s1}
\end{equation}
This was used as the loss function to train the PQANet.

\subsection{Stage 2: Spatio-temporal Pooling}

\noindent\textbf{Network Architecture.} Most of the existing spatial-temporal pooling approaches in deep VQA only take the obtained patch-level quality indices as inputs \cite{C3DVQA, Kim2018DeepVQ,E2E}, but ignore the information within the distorted content. In this work, inspired by \cite{wu2022discovqa} in which only temporal information was extracted for pooling, we designed a novel spatial-temporal aggregation network, shown in Figure \ref{fig:framework}, which accepts both patch-level score tensors and the distorted features maps (extracted in Stage 1) to obtain the final sequence level quality score. 

First, the network takes all the patch level quality indices generated in Stage 1, groups them in a 3D tensor and normalizes the tensor size to $W_\mathcal{P}\times H_\mathcal{P} \times T_\mathcal{P}$ (the re-sampling is based on simple local mean). Here we set up $W_\mathcal{P}=16$, $H_\mathcal{P}=9$ and $T_\mathcal{P}=10$. The network also re-uses the feature maps $\mathcal{F}_D$ (Level 6 and Level 3 only \cite{li2021efficient,li2019frd})  generated by the FE module in Stage 1 based on all the distorted patches, and combines these feature maps in a 3D tensor and re-sample it to the same resolution of the patch quality tensor $16\times 9 \times 10$ (each component in this tensor corresponds to two feature maps at level 6 and 3).

The distorted feature tensor is then fed into two 3D convolution blocks (each containing two 3D CNN layers with kernel size of $3\times 3\times 3$,  channel number of 8, and stride of 1)  followed by a Softmax function to capture spatio-temporal information, which is used to weight the patch level quality indices. The weighted indices are then combined to produce the final quality score for this video sequence.

\noindent\textbf{Training Material.} In Stage 2, we use video quality databases containing subjective ground truth labels to train the spatio-temporal aggregation network. As we still adopt a ranking-inspired training strategy (as discussed below), we are able to combine training content from multiple video quality databases for the first time, which significantly improves the training effectiveness. We employed two databases, VMAF+~\cite{w:VMAF} and IVP \cite{w:IVP} in the Stage 2 training.  VMAF+ was used to train the original VMAF model, while IVP contains diverse distortion types generated by different compression algorithms.

Each sequence in both databases was first segmented into non-overlapped 256$\times$256$\times$3$\times$12 patches and fed into the PQANet to obtain patch level quality indices. These quality indices together with their corresponding sequence level subjective opinion score (as the training target) were used as the training material in Stage 2.

\noindent\textbf{Training Strategy.} As in Stage 1, a similar ranking-inspired training strategy was also employed to train the spatio-temporal aggregation network (STANet). For each pair of randomly selected distorted sequences (from the same database), denoted as $X$ and $Y$, given their the input patch quality indices, $\{Q_{\mathcal{P}_{X,i}}\}$ and $\{Q_{\mathcal{P}_{Y,i}}\}$, their patch-level features $\{\mathcal{F}_{D_{X,i}}\}$ and $\{\mathcal{F}_{D_{Y,i}}\}$, and the corresponding normalized ([0, 100]) subjective opinion scores (e.g., MOS or 100-DMOS), $S_X$ and $S_Y$, we calculate the following loss to optimize STANet.
\begin{small}
\begin{gather}
\footnotesize
\resizebox{0.98\hsize}{!}{$\delta = \mathrm{STANet}(\{Q_{\mathcal{P}_{X,i}}\}, \{\mathcal{F}_{D_{X,i}}\})-\mathrm{STANet}(\{Q_{\mathcal{P}_{Y,i}}\}, \{\mathcal{F}_{D_{Y,i}}\})$} \\
\mathcal{L}_{s2} = \parallel \delta-(S_X-S_Y)\parallel_2.
\label{eq:BCEloss}
\end{gather}
\end{small}

We have used a total number of 16,000 sequences pairs from both databases in the Stage 2 training. It is noted that we take the subjective score difference into account rather than perform the binarization as in Stage 1. This is because we believe that this difference value (from real ground truth)  provides more important and accurate information compared to that from VMAF.  

\section{Experiment Setup}
\label{sec:configuration}

\noindent\textbf{Implementation Details.} Pytorch 1.10 was used to implement both networks, with the following training parameters: Adam optimization \cite{kingma2014adam} with $\beta_1$=0.9 and $\beta_2$=0.999; 60 training epochs; batch size of 4; the initial learning rate is 0.0001 with a weight decay of 0.1 after every 20 epochs. Both training and evaluation were executed on a computer with a 2.4GHz Intel CPU and an NVIDIA P100 GPU.\\
\noindent\textbf{Evaluation Dataset and Configuration.} To evaluate the  performance of the proposed methods and their model generalization, eight commonly used video quality datasets\footnote{In this paper, our evaluation solely focuses on HD content, as this is the dominant format in modern video streaming.} were employed including NFLX \cite{w:VMAF}, NFLX-P \cite{w:VMAF}, BVI-HD \cite{j:Zhang7},  BVI-CCHD \cite{j:Zhang16}, BVI-CCHDDO \cite{c:Zhang24}, MCL-V \cite{j:Lin4}, SHVC \cite{r:JCTVCW0095} and VQEGHD3 \cite{r:vqegHD}. In this experiment, we did not re-train our model (or other learning-based methods) on these databases and test their (intra-database) performance through cross-validation, as this often leads to less meaningful results (with correlation coefficients that are close to 1) due to overfitting issues \cite{Kim2018DeepVQ,C3DVQA,E2E}. Instead, we only used the training databases mentioned in Section \ref{sec:Proposed Approach} to optimize our approaches, and fixed the model parameters during evaluation. This experimental design challenges against the generalization of the tested VQA methods.

In the evaluation stage, for FR-RankDVQA, each distorted sequence and its corresponding reference are segmented into non-overlapping 256$\times$256$\times$12 spatio-temporal patches and converted into YCbCr 4:4:4 formats as the input of the PQANet. The output quality indices for all these patches are then employed as the input of the STANet to obtain the final quality index of this sequence. For NR-RankDVQA, only distorted sequence is required and it is processed following the same approach described above.


\noindent\textbf{Evaluation metrics.} To assess the performance of each VQA method, four correlation statistics have been calculated: the Spearman Rank Order Correlation Coefficient (SROCC), the Pearson Linear Correlation Coefficient (PLCC), the Outlier ratio (OR) and the Root
Mean Squared Error (RMSE) \cite{r:vqeg} to appraise prediction accuracy (PLCC, RMSE), monotonicity (SROCC) and consistency (OR). Due to the limited space in the main paper, we only present SROCC results here, and include results based on the other three metrics in the \textit{Supplementary Material}. Additionally, a significance test was also performed to differentiate between the proposed methods (FR- and NR-RankDVQA) and their benchmarks in each category (FR and NR) on each test database. The approach in \cite{j:LIVE,j:movie} was used in which an F-test was conducted on the residuals between the subjective opinion scores and the predicated quality scores by the tested VQA methods through a non-linear regression employing a logistic function \cite{r:vqeg}.

\begin{table*}[ht]
	\scriptsize
	\centering
 \begin{tabular}{l||llllllll|l|l}
		\toprule
		\centering
        \setlength{\tabcolsep}{12pt}
		SROCC$\uparrow$ (F-test) & NFLX & NFLX-P & BVI-HD & BVI-CCHD & BVI-CCHDDO & MCL-V & SHVC  & VQEGHD3 & {Overall} & Overall (SS)\\
		\midrule \midrule
              \multicolumn{11}{c}{Full Reference VQA Methods}\\
              \midrule
		PSNR  & 0.6218 (-1) & 0.6596 (-1) & 0.6143 (-1) & 0.6166 (-1) & 0.7497 (-1) & 0.4640 (-1) & 0.7380 (-1)  & 0.7518 (-1) & 0.6520 & 0.9476\\\midrule
		SSIM \cite{ssim} & 0.5638 (-1) & 0.6054 (-1) & 0.5992 (-1) & 0.7194 (-1) & 0.8026 (-1) & 0.4018 (-1) & 0.5446 (-1)  & 0.7361 (-1) & 0.6216 &0.9451 \\\midrule
		MS-SSIM \cite{c:mssim} & 0.7136 (-1) & 0.7394 (-1) & 0.7652 (-1) & 0.7534 (-1) & 0.8321 (0) & 0.6306 (-1) & 0.8007 (0)  & \textcolor{blue}{0.8457} (0) & 0.7601 &\textcolor{blue}{0.9477}\\\midrule
		DeepQA \cite{Kim_2017_CVPR} & 0.7298 (-1) & 0.6995 (-1) & 0.7106 (-1) & 0.6202 (-1) & 0.6705 (-1) & 0.6832 (-1) & 0.7176 (-1)  & 0.7881 (-1) & 0.7024& 0.8854\\\midrule
		LPIPS \cite{Lpips}&0.6793(-1) &	0.7859 (-1)	& 0.6670 (-1)& 	0.6838 (-1)&	0.7678 (-1)&	0.6579 (-1)& 0.6360 (-1)	&0.8075 (0) & 0.7107& 0.9041\\\midrule	
        DeepVQA \cite{Kim2018DeepVQ} & 0.7352 (-1) & 0.7609 (-1) & 0.7330 (-1) & 0.6924 (-1) & 0.8120 (0) & 0.6142 (-1) & 0.8041 (0)  & 0.7805 (-1) & 0.7540& 0.9060\\\midrule
        C3DVQA \cite{C3DVQA} & 0.7730 (-1) & 0.7714 (-1) & 0.7393 (-1) & 0.7203 (-1) & 0.8137 (0) & 0.7126 (-1) & 0.8194 (0)  & 0.7329 (-1) & 0.7641& 0.9421\\\midrule
	    DISTS \cite{DISTS} & 0.7787 (-1) & \textcolor{red}{0.9325} (0) & 0.7030 (-1) & 0.6303 (-1) & 0.7442 (-1) & \textcolor{blue}{0.7792} (-1) & 0.7813 (0) & 0.8254 (0) &  0.7718   & 0.9235\\\midrule
        ST-GREED \cite{STGREED}& 0.7470 (-1) & 0.7445 (-1) & 0.7769 (-1) & 0.7738 (-1) & 0.8259 (0) & 0.7226 (-1) & 0.7946 (0)  & 0.8079 (0) & 0.7842 &0.9460\\\midrule


		VMAF 0.6.1 \cite{w:VMAF}   & \textcolor{blue}{0.9254} (0)   & 0.9104 (0)  & \textcolor{blue}{0.7962} (-1)& \textcolor{blue}{0.8723} (0)& \textcolor{blue}{0.8783} (0)& 0.7766 (-1)&    \textcolor{blue}{0.9114} (0)   & 0.8442 (0) & \textcolor{blue}{0.8644}& 0.9455\\\midrule 
     
        \textbf{FR-RankDVQA}    & \textcolor{red}{0.9393}  & \textcolor{blue}{0.9184}  & \textcolor{red}{0.8659}  & \textcolor{red}{0.8991}  & \textcolor{red}{0.9037}   & \textcolor{red}{0.8391}   & \textcolor{red}{0.9142} & \textcolor{red}{0.8979}  & \textcolor{red}{\textbf{0.8972}}&\textcolor{red}{\textbf{0.9814}}\\
            \midrule\midrule
            \multicolumn{11}{c}{No reference VQA methods}\\\midrule
		VIIDEO \cite{VIIDEO} & 0.4550 (-1) & 0.5527 (-1) & 0.1297 (-1) & 0.1308 (-1) & 0.2523 (-1) & 0.0406 (-1) & 0.2033 (-1)  & 0.1881 (-1)  & 0.2440  &0.3087\\\midrule
		TLVQM \cite{TLVQM} & 0.4652 (-1) & 0.4720 (-1) & 0.3124 (-1) & 0.1622 (-1) & 0.3420 (-1) & 0.2758 (-1) & 0.4983 (0)  & 0.5382 (0)  & 0.3469 & 0.8239\\\midrule
        BRISQUE \cite{brisque}  & 0.7828 (0) & 0.7861 (0) & 0.2033 (-1) & 0.3738 (-1) & 0.3746 (-1) & 0.3154 (-1) & 0.3601 (-1) & 0.5467 (0) & 0.4716 & 0.5894\\\midrule
  	NIQE \cite{niqe}  & 0.7959 (0) & \textcolor{red}{0.8269} (0) & 0.1932 (-1) & 0.4247 (-1) & 0.5225 (-1) & 0.3985 (-1) & 0.6210 (0)&         0.5291 (0)& 0.5390 & 0.7029\\\midrule
        MDTVSFA \cite{MDTVSFA} & 0.5137 (-1) & 0.6024 (-1) & 0.3725 (-1) & 0.4068 (-1) & 0.5547 (-1) & 0.5712(0) & 0.6165 (0)  & 0.6422 (0)  & 0.5311 &0.8872\\\midrule
        CONVIQT\cite{conviqt} & 0.6989 (-1) &\textcolor{blue}{0.7962} (0)	& 0.3489 (-1) &	0.3706 (-1)	& 0.5381(-1) &	0.6323 (0)	& 0.4983 (0)	& 0.6217 (0) &0.5631 & 0.6846\\\midrule
        VIDEVAL \cite{VIDEVAL} & 0.7899 (0) & 0.7261 (0) & \textcolor{blue}{0.5884} (-1) & 0.6974 (0) & 0.7620 (0) & 0.4836 (-1) & 0.6428 (0) & 0.5326 (0) & 0.6529 & 0.8621 \\\midrule
		GSTVQA \cite{GSTVQA}&\textcolor{blue}{0.8109} (0) & 0.7858 (0) & 0.4132 (-1) & \textcolor{blue}{0.7447} (0) & \textcolor{blue}{0.7665} (0) & \textcolor{blue}{0.7385} (0) & \textcolor{blue}{0.6710} (0) & \textcolor{blue}{0.7011} (0) & \textcolor{blue}{0.7040} &\textcolor{blue}{0.9014}\\\midrule
		\textbf{NR-RankDVQA}    & \textcolor{red}{0.8346}   & \textcolor{blue}{0.7944}   & \textcolor{red}{0.7326} & \textcolor{red}{0.7628} & \textcolor{red}{0.7994} & \textcolor{red}{0.7631} &    \textcolor{red}{0.7118}   & \textcolor{red}{0.8346}   & \textcolor{red}{\textbf{0.7791}} & \textcolor{red}{\textbf{0.9266}}\\
                \midrule\midrule
                \multicolumn{11}{c}{Ablation Study Results}\\\midrule

        {V1 ($\ell 1$)}  &  0.8793(0) & 0.8816 (0) & 0.7583 (-1) & 0.7792 (-1) & 0.8523 (0) & 0.7678 (-1) & 0.8238 (0)  & 0.8501 (0) & 0.8190 & 0.9417 \\\midrule
         {V2 ($\ell 2$)}    & 0.8812 (0)  & 0.8883 (0)  & 0.7612 (-1) & 0.7794 (-1) & 0.8568 (0) & 0.7696 (-1) & 0.8234 (0) &0.8507 (0) &0.8263 &  0.9467 \\\midrule

         V3 (C3D)   & 0.9034(0)  & 0.8964 (0)  & 0.8233 (0)  & 0.8763 (0)  & 0.8961 (0)  & 0.8054 (0)  & 0.8692 (0) & 0.8465 (0) & 0.8653  & 0.9693\\\midrule
         {V4 (S1)}    & 0.9201(0)  & 0.8983 (0)  & 0.8361 (0)  & 0.8825 (0)  & 0.8987 (0)  & 0.8231 (0)  & 0.8966 (0) & 0.8544 (0) & 0.8762 & 0.9617\\\midrule
          \textbf{FR-RankDVQA}    & \textcolor{red}{0.9393}  & \textcolor{red}{0.9184}  & \textcolor{red}{0.8659}  & \textcolor{red}{0.8991}  & \textcolor{red}{0.9037}   & \textcolor{red}{0.8391}   & \textcolor{red}{0.9142} & \textcolor{red}{0.8979}  & \textcolor{red}{\textbf{0.8972}}&\textcolor{red}{\textbf{0.9814}}\\
		\bottomrule
        \end{tabular}
        \vspace{0.5em}
        
        \caption{Performance of the proposed methods, other benchmark approaches and ablation study variants on eight HD test databases. The values in each cell x(y) correspond to the SROCC value (x) and F-test result (y) at 95\% confidence interval. y=1 suggests that the metric is superior to FR-RankDVQA in the full reference track or NF-RankDVQA in the no reference track (y=-1 if the opposite is true), while y=0 indicates that there is no significant difference between them. The figures in \textcolor{red}{red} and \textcolor{blue}{blue} indicate the highest and second highest SROCC values respectively in each column.}
        \vspace{-1pt}
	\label{tab:results}
\end{table*}

\noindent\textbf{Benchmark VQA Methods.} 
To benchmark the performance of the proposed  methods, we tested ten full reference quality assessment methods, including three conventional ones: PSNR, SSIM \cite{ssim}, and MS-SSIM \cite{c:mssim}; and five deep VQA methods:\footnote{The selection of benchmark deep VQA methods are based on the reported performance in the original literature and the availability of their pre-trained models.} DeepQA \cite{Kim_2017_CVPR}, DeepVQA \cite{Kim2018DeepVQ}, C3DVQA \cite{C3DVQA}, DISTS \cite{DISTS} and LPIPS \cite{Lpips}; and two machine learning (SVM regression) based quality metrics: GREED \cite{STGREED} and VMAF \cite{w:VMAF} We have also evaluated eight NR VQA methods, among which MDTVSFA \cite{MDTVSFA}, CONVIQT\cite{conviqt} (unsupervised learning), VIDEVAL \cite{VIDEVAL} GSTVQA \cite{GSTVQA} are deep learning based methods, while VIIDEO \cite{VIIDEO}, BRISQUE \cite{brisque}, NIQE \cite{niqe} and TLVQM \cite{TLVQM} are classic or machine learning-based approaches. For all the deep learning-based quality metrics, we used their publicly available pre-trained models for result generation\footnote{As the primary contribution of this work is to develop generic deep VQA methods which does not require cross-validation, we did not re-training benchmark methods on different databases. The performance presented here shows the generalization characteristic of each model.}.

\section{Experiments}
\label{sec:results}

\begin{figure*}[ht]
\centering
\includegraphics[width=0.88\linewidth]{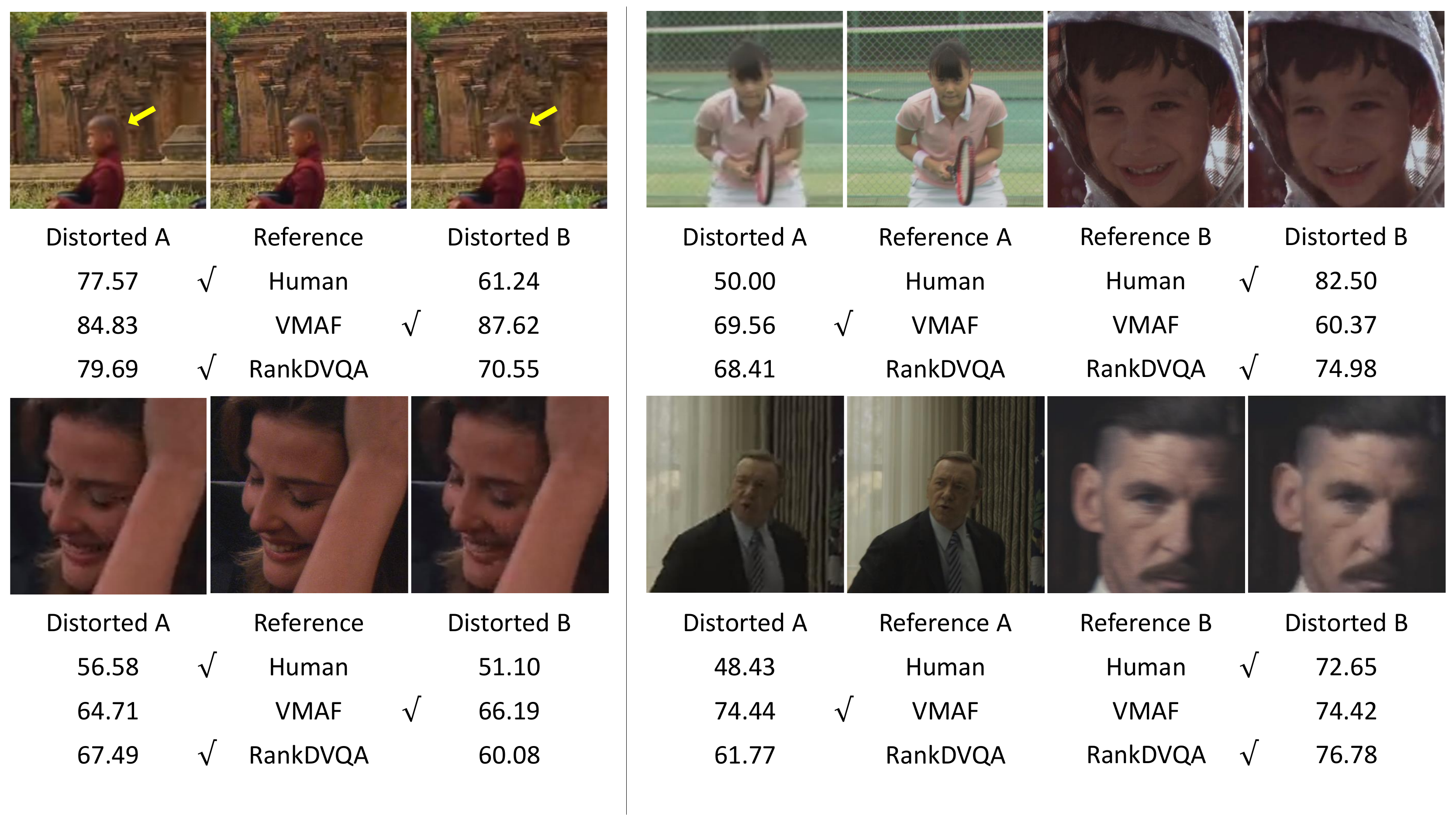}
\vspace{-0.8em}
\caption{Visual examples demonstrating the superiority of the proposed FR quality metric. For the frames on the left, it can be seen that the distorted video A is given a higher subjective quality score than video B, and FR-RankDVQA correctly predicts this. Similarly for frames on the right, although distorted video B shows less visual distortion from its reference compared to distortion between distorted video A and reference A, VMAF fails to predict the correct quality rank (the proposed method does). \label{fig:Subjective}}

\vspace{-10pt}
\end{figure*}


\subsection{Quantitative Evaluation}
\label{sec:quantitative}


\noindent\textbf{Full Reference VQA Method Comparison.} The results in Table \ref{tab:results} shows that FR-RankDVQA achieves the highest overall SROCC value of 0.8972 among all tested FR quality metrics across eight databases. The second best performer is VMAF, which is consistently inferior to FR-RankDVQA. Based on the F-test results, their performance difference is significant on BVI-HD and MCL-V databases. It is also noted that all the deep VQA methods underperform VMAF, and none of them achieve a SROCC value higher than 0.8. 

In terms of complexity, the runtime of FR-RankDVQA is 3.88 times slower than VMAF,  similar to other under-performing deep FR-VQA methods, such as DeepVQA \cite{Kim2018DeepVQ} (4.05$\times$) and C3DVQA \cite{C3DVQA} (3.37$\times$). Comprehensive complexity figures can be found in the  \textit{Supplementary Material}.

\noindent\textbf{No Reference VQA Method Comparison.} In the no reference case, NR-RankDVQA also offers the best overall performance compared to other tested NR VQA methods, with an average SROCC of 0.7791. This figure is much higher than that of the second best quality metric, GSTVQA \cite{GSTVQA} (SROCC = 0.7040), and than some of the full reference VQA methods including PSNR, SSIM, MS-SSIM, WaDIQA, DeepQA, DeepVQA and C3DVQA.

\noindent\textbf{Assessing Content from Single Sources.} As pointed out in Section \ref{sec:stage1}, it is important for a VQA method to accurately differentiate the quality between distorted versions from the same source or from different sources. The latter has been effectively evaluated above, when the SROCC values were calculated for the whole database. In order to test the single source case, following the evaluation procedure in \cite{hou2022perceptual}, we first compute the SROCC value for all the distorted sequences from each individual source, and average them among all sources within each database. The overall SROCC values (for all test databases) for selected VQA methods are summarized in Table \ref{tab:results} (Overall (SS) column), while comprehensive results can be found in \textit{Supplementary Material}. It can be observed that the proposed FR- and NR-RankDVQA also outperform the benchmark algorithms in each track (FR and NR). Both properties of the proposed method (full reference) have further been confirmed in Figure \ref{fig:Subjective} when compared with VMAF.

\subsection{Ablation Study}

To evaluate the primary contributions of this work, we have evaluated the following RankDVQA variants in an ablation study. Here we only focus on the FR scenario.

\noindent\textbf{Effectiveness of the ranking-inspired Losses.} One major novelty of this work is the Ranking-inspired training strategy. To evaluate its contribution and compare it with commonly used loss functions, $\ell 1$ and $\ell 2$ (between the output of the network and the training target), in the VQA research community, we have implemented two variants V1($\ell 1$) and V2($\ell 2$) which replaced the Ranking-inspired loss functions with $\ell 1$ or $\ell 2$ in both stages. The results for both variants are also shown in Table \ref{tab:results}, which are also outperformed by FR-RanDVQA on all tested databases.

\noindent\textbf{Effectiveness of PQANet.}
In order to validate the contribution of PQANet to the overall design, we evaluate the training framework on a different deep VQA model, C3DVQA, creating V3(C3D). Except for the exchange of the first-stage network, all other training settings (including the second stage) are unchanged. It can be observed in Table \ref{tab:results} that V3(C3D) is also outperformed by FR-RankDVQA, which implies the effectiveness of the PQANet.

\noindent\textbf{Effectiveness of STANet.} To verify the effectiveness of the STANet, we have replaced this aggregation network with a simple arithmetic average operation, which is typically used for spatio-temporal pooling. Its performance is denoted as V4(S1) in Table \ref{tab:results}. It can be observed by comparing this variant with FR-RankDVQA, that the latter offers higher SROCC values on all tested video databases.  

\section{Conclusion}
\label{sec:conclusion}
 In this paper, we propose new deep VQA methods based on a ranking-inspired hybrid training methodology. We introduce for the first time, the use of a large scale training database (created without the need to perform expensive and time consuming subjective tests) to optimize deep networks with high capacity for quality assessment. This also supports the  combination of multiple existing video quality databases (with subjective ground truth labels) to train an aggregation network for spatio-temporal pooling. 
 
 The proposed methods, FR-RankDVQA and NR-RankDVQA, were fully tested on eight evaluation databases, and were shown to exhibit higher correlation with opinion scores when compared to other full reference and no reference VQA methods. To the best of our knowledge, FR-RankDVQA is the first deep VQA method that consistently outperforms VMAF on multiple databases without conducting intra-database cross validation. Future work should focus on more sophisticated pooling network architectures and complexity reduction. 

\vspace{5pt}

\noindent\textbf{Acknowledgement.} The authors appreciate the funding from the UKRI MyWorld Strength in Places Programme (SIPF00006/1), the University of Bristol, and the China Scholarship Council.

{\small
\bibliographystyle{ieee_fullname}
\bibliography{egbib}
}

\end{document}